\documentclass[10pt, twosided]{article}
\date{}
\RequirePackage{titlesec}
\RequirePackage{amsmath,esdiff}
\RequirePackage{amssymb}
\RequirePackage{authblk}

\RequirePackage[normalem]{ulem}
\RequirePackage{bbold,bm}
\RequirePackage{BOONDOX-cal}
\usepackage{braket}
\usepackage{float}
\RequirePackage[colorlinks=true,linkcolor=blue,citecolor=red,urlcolor=red]{hyperref}
\RequirePackage{geometry}
\RequirePackage{setspace}
\RequirePackage{cite}
\RequirePackage{flushend}
\RequirePackage{mathrsfs}
\RequirePackage{hyperref}
\RequirePackage{graphicx}
\RequirePackage{cancel}
\RequirePackage{array,esint}
\RequirePackage[most]{tcolorbox}
\graphicspath{}
\titleformat{\section}{\fontsize{12}{12}\bfseries}{\thesection}{1em}{}

\begin{document}
\title{{\bf{\textbf{Quantum harmonic oscillator in a time dependent noncommutative background}
}}}
\author{
{\bf {\normalsize Manjari Dutta}$^{a}
$\thanks{manjaridutta@boson.bose.res.in, chandromouli15@gmail.com}},
{\bf {\normalsize Shreemoyee Ganguly}
$^{b}$\thanks{ganguly.shreemoyee@gmail.com}},
{\bf {\normalsize Sunandan Gangopadhyay}
$^{c}$\thanks{ sunandan.gangopadhyay@bose.res.in, sunandan.gangopadhyay@gmail.com}}
\\
$^{a,c}$ {\normalsize Department of Astrophysics and High Energy physics},\\
{\normalsize S.N. Bose National Centre for Basic Sciences},\\
{\normalsize JD Block, Sector III, Salt Lake, Kolkata 700106, India}\\
$^{b}$ {\normalsize Department of Basic Science and Humanities,}\\
{\normalsize University of Engineering and Management (UEM),}\\
{\normalsize B/5, Plot No.III, Action Area-III, Newtown, Kolkata 700156}
}

\maketitle

\begin{abstract}
\noindent This work explores the behaviour of a noncommutative harmonic oscillator in a time-dependent background, as previously investigated in \cite{Dey}. Specifically, we examine the system when expressed in terms of commutative variables, utilizing a generalized form of the standard Bopp-shift relations recently introduced in \cite{spb}. We solved the time dependent system and obtained the analytical form of the eigenfunction using the method of Lewis invariants, which is associated with the Ermakov-Pinney equation, a non-linear differential equation. We then obtain exact analytical solution set for the Ermakov-Pinney equation. With these solutions in place, we move on to compute the dynamics of the energy expectation value analytically and explore their graphical representations for various solution sets of the Ermakov-Pinney equation, associated with a particular choice of quantum number. Finally, we determined the generalized form of the uncertainty equality relations among the operators for both commutative and noncommutative cases. Expectedly, our study is consistent with the findings in \cite{Dey}, specifically in a particular limit where the coordinate mapping relations reduce to the standard Bopp-shift relations.

\end{abstract}
\newpage
\section{Introduction}
The concept of noncommutative (NC) spacetime as well as the discreteness of spacetime structure seems to be an essential requirement to unify the general theory of relativity and the laws of quantum mechanics. When measuring the small distance at around Planck scale, both the momentum and the energy of the probing particle become very high, in accordance with Heisenberg's uncertainty principle \cite{hei1}. 
Consequently, this leads to the collapse of the spacetime manifold.\,This collapse ultimately causes the breakdown of the general theory of relativity as well as the laws of gravitation at this extremely small length scale. Therefore, in order to unify both the general theory of relativity and quantum field theory, it is necessary to employ a minimum observable length, beyond which measurement is impossible in the spacetime manifold.\\
\noindent The idea of NC quantum theory was first historically introduced by Heisenberg (in the late 1930, he wrote to Pierls in a letter \cite{hei}). Finally, the idea is formalised in the pioneering work~\cite{Synder} by Snyder and
the study of quantum mechanical systems in NC spaces has captivated theoretical physicists. In that work, by providing an example of a Lorentz invariant discrete spacetime, it is shown that the usual assumption of continuum spacetime is not required by Lorentz invariance. Immediately after this work, an extension to this treatment is also performed by \cite{Yang} for the case of curved spacetime (specifically de Sitter space). Although, at that time that idea is largely ignored,
the NC nature of spacetime is later strongly supported by string theory \cite{amati, gross, vene} which is one of the current theories of quantum gravity. The theory comes up with a concept of a finite length $l_s$ which is actually the length of the strings and measures the short distance structure. Hence, it is not possible to observe the distance which is even smaller than $l_s$. It was also shown in the seminal work \cite{sw} that at a certain low energy limit, string theory can be treated as an effective quantum field theory in NC spacetime \cite{Doug, con}. However, this is a fact that, almost all the theories of quantum gravity imply the existence of minimum measurable length as well as the discreteness of spacetime which can resist the gravitational collapse at microscopic level. Hence, the consideration of NC spacetime can also ensure gravitational stability~\cite{Dop1, Dop2}. The effects of NC spacetime has been studied in single field inflation in \cite{bal, nau}. In particular, the two and three point correlation functions for the curvature perturbation in Friedmann-Lemaître-Robertson-Walker spacetime were computed. It was found that the power spectrum was anisotropic due to the effects of NC geometry. It should also be noted that the Lorentz symmetry in NC spacetime is violated due to the presence of a fundamental length scale. This violation of Lorentz symmetry and hence Poincaré symmetry also comes with the violation of the CPT theorem, microcausality and spin statistics. The Poincaré symmetry can be restored by the introduction of what is known as a deformed coproduct in the literature \cite{chai, bloh, dimi, oeck}. This deformed coproduct has been used extensively in relativistic quantum field theories \cite{bal1} and its applications have been studied in \cite{bal2, bwc}. Apart from these, the problem of understanding quantum spacetime is also approached by loop quantum gravity \cite{ast, rov}.
As a result, several studies on quantum mechanical systems in NC spaces have been conducted and reported in the literature~\cite{suss}-\cite{fgs}. 

\noindent The simplest model of  NC spacetime is a two-dimensional quantum mechanical space, where the standard set of commutation relations between the canonical coordinates is replaced by NC commutation relations $[X, Y]=i\theta$, where $\theta$ is a positive real constant. Studies involving NC spaces may also include the NC commutation relations $[P_x, P_y]=i\Omega$, where $\Omega$ is a real constant. However, more interesting structural form of noncommutativity can be found as the function of momenta and coordinates \cite{kem1, kem2, fr1, fr2, fr3, fr4} in the literature. Explicit time dependence of NC parameters can also be introduced. Quantum theories on such spaces have been studied in \cite{Dey, SG1, SG2, SG3}. The motivation for such studies was that very little was known about these models. The physical consequences that would be implied by such theories were investigated in \cite{Dey}. In the realm of noncommutativity, it has been found to be convenient to use the standard Bopp-shift \cite{mez} relations to map a system in NC space in terms of commutative variables. However, in a recent communication \cite{spb}, a modified version of the standard Bopp-shift relations have been proposed. The main difference between two types of coordinate mapping relations lies in the introduced relations in \cite{spb}, where the commutation relations between the spatial coordinates and the momentum coordinates simply obey the phase-space commutation relation in ordinary quantum mechanics, that is, $[X_j,P_k]=i\,\delta_{jk}$\,; where $j\,,k\,=\,1,2$. This contrasts with the standard Bopp-shift relations, as the commutator $[X_j,P_k]$ is a function of the NC parameters $\theta$ and $\Omega$. Another distinction arises from the fact that, although in both the standard version and the modified version of the coordinate mapping relations, the NC parameters $\theta, \Omega$ are real constants, the second version imposes an additional condition $\theta\Omega\,<\,0$ to prevent the appearance of imaginary numbers in the structure of the coordinate mapping relations. Moreover, the most interesting feature, as also observed in \cite{spb}, brought about by this newly introduced noncommutativity is its ability to include a time reversal (TR) symmetry breaking scale invariant term in the latter form of the Hamiltonian, expressed in terms of commutative variables. Under adiabatic conditions, the presence of such a scale invariant term in the Hamiltonian indicates the potential emergence of a non-trivial Berry's geometric phase \cite{berry1, berry2} when the Hamiltonian evolves periodically with respect to time. Due to these interesting features, this modified NC structure invites further exploration, particularly in the context of simple prototype models within its framework. \\
\noindent In contrast to the study mentioned above, where the NC parameters were set to be constants, we adopt their coordinate mapping relations by allowing the NC parameters to vary with respect to time in our work. In fact, as discussed in several sources, including \cite{reuter}-\cite{reuter-book}, the time varying NC parameters can be thought to arise from the renormalization group flow of Newton's gravitational constant, with the energy scale being inversely proportional to the cosmic time. Since, the minimum measurable length is explicitly linked to Newton's gravitational constant, it can be assumed to vary with respect to time, just as the Newton's constant does. In literature, there are already some studies on the exact quantum theory in the context of time dependent NC space.
In \cite{strek2}, which is a sequel to \cite{strek1}, the exact propagator of a two dimensional harmonic oscillator in time dependent NC space was found.
The simplest prototype model of a harmonic oscillator in such time-dependent NC background was first discussed in \cite{Dey} where the time dependent model is exactly solved using Lewis method of invariant \cite{Lewis, Lewis2, Lewis3}. As the method of invariant as well as the Lewis invariant is associated with a non linear differential equation, known as Ermakov-Pinney equation \cite{Erm}-\cite{Pin} in the literature, the equation behaves like a constraint relation to be obeyed by the parameters those form the eigenfunction of the Hamiltonian. However, under a certain integrability condition, it is also possible to deduce a few sets of explicit functional forms for the parameters constrained by EP equation.\,Thus, every solution set of the EP equation can provide an explicit form for the eigenfunction of the Hamiltonian, and a class of exact solutions for the Hamiltonian can be obtained using this technique. The explicit solutions found in \cite{Dey} allowed for an analysis of the uncertainty relations. The study revealed that the lower bounds became time-dependent functions. The uncertainties in the Glauber coherent states, constructed in the standard way by the action of the displacement operator, were also found. 
Later, with this time dependent noncommutativity in place, there were several works \cite{SG1}-\cite{SG3} on the model of exactly solvable time dependent Hamiltonians. In \cite{SG1} and \cite{SG2}, using Lewis technique, a class of exact solutions are obtained for a damped harmonic oscillator both in the absence and presence of an external magnetic field and placed in the time dependent NC space. In another recent communication \cite{SG3}, we have used the new change of variables (discussed above) \cite{spb} to study the explicit existence of Berry's geometric phase \cite{berry1, berry2} formed in time dependent NC harmonic oscillator systems. Hence, all those studies \cite{Dey, SG1, SG2, SG3} concerning the exactly solvable models in time dependent NC space, suggest a possibility of developing a quantum theory in the background of time dependent NC space. It is highly desirable to study various models on different types of NC backgrounds. Such studies would definitely enrich our understanding. The first thing to look for in these studies is whether the models on different time dependent NC backgrounds still allow for explicit solvability. These are some of the main motivations for our present study. \\
\noindent In the context of the project mentioned above, the main purpose of the current study is to initiate a quantum theory within a time dependent NC framework which adheres to the NC algebra proposed by \cite{spb}. For doing so, we examine a time independent harmonic oscillator situated in a time dependent NC background utilizing coordinate mapping relations introduced by \cite{spb}. We followed the approach taken by \cite{Dey} where standard Bopp-shift relations were used to map the original Hamiltonian of the time-independent harmonic oscillator in NC space into ordinary commutative variables. \\
\noindent The organization of our work is as follows. In section 2, we consider a two-dimensional, time independent quantum harmonic oscillator in time-dependent NC space associated with a modified version of Bopp-shift relations. We construct the original Hamiltonian in NC space and then express it in terms of standard commutative variables, resulting in a Hamiltonian containing a Zeeman term and a scale-invariant term. In section 3, we use the Lewis-Riesenfeld \cite{Lewis} approach to solve the resultant time-dependent Hamiltonian, referring to one of our recent communications \cite{SG3} that addresses a similar time-dependent system using the same approach. We begin by presenting the form of the Lewis invariant associated with the non-linear Ermakov-Pinney (EP) \cite{Erm}-\cite{Pin} equation. Then, we introduce the eigenfunction of the Hamiltonian, which is a product of the eigenfunction of the Lewis invariant and a time-dependent phase factor, also known as the Lewis phase in the literature. In section 4, we expand upon the EP solution set provided in \cite{Dey} to solve the non-linear EP equation consistent with the Chiellini integrability condition \cite{chill}. In Section 5, we devise a procedure to calculate the expectation value of the Hamiltonian with respect to its own eigenstate. Using this expression, we study the evolution of the energy expectation value of the system with time for various types of EP solution set, both analytically and graphically. We also derive the generalized form of the uncertainty equality relations obeyed by the commutative and NC coordinate operators. Finally, in Section 6, we summarize our results.
\section{Two dimensional harmonic oscillator model in modified noncommutative space} 
\noindent We start by looking at a harmonic oscillator in a NC space, incorporating a recent modification discussed in \cite{spb}. The Hamiltonian in the NC space shares a similar structure with that examined in \cite{Dey}.
\noindent The Hamiltonian of the system in NC space reads,
\begin{equation}
H(t)|_{NC}=\dfrac{1}{2\,M}(P_1^2+P_2^2)+\dfrac{1}{2}\,M\,\omega^2(X_1^2+X_2^2)~;\label{Ham2}
\end{equation}
where $M$ and $\omega$ are the constant mass and angular frequency of the harmonic oscillator respectively. The commutation relations among the NC variables which are position and momentum 
coordinates $(X_i,P_i)$, are as follows (considering, $\hbar=1$) 
\begin{align}
[X_1,X_2]=i\theta(t)~,~[P_1,P_2]=i\Omega(t)~,~
[X_1,P_1]=i=[X_2,P_2]~;
\end{align}
where $\theta(t)$ and $\Omega(t)$ denote the NC parameters for space 
and momentum respectively. The commutation relations among the canonical variables $(x_i,p_i)$ in commutative space are
\begin{equation}
[x_i,p_j]=i\delta_{i,j}, [x_i,x_j]=0=[p_i,p_j]; (i,j=1,2).
\end{equation}
In order to represent the Hamiltonian [eqn.(\ref{Ham2})] in commutative space, we apply the modified transformation relations between the NC and commutative coordinates which read \cite{spb},  
\begin{eqnarray}
&X_1=x_1-\dfrac{\theta(t)}{2}p_2+\dfrac{\sqrt{-\theta(t)\Omega(t)}}{2}x_2~,~X_2=x_2+\dfrac{\theta(t)}{2}p_1-\dfrac{\sqrt{-\theta(t)\Omega(t)}}{2}x_1~;\label{Xbop}\\
&P_1=p_1+\dfrac{\Omega(t)}{2}x_2+\dfrac{\sqrt{-\theta(t)\Omega(t)}}{2}p_2~,~P_2=p_2-\dfrac{\Omega(t)}{2}x_1-\dfrac{\sqrt{-\theta(t)\Omega(t)}}{2}p_1~;\label{Pbop}
\end{eqnarray}
where $\theta(t)\Omega(t)<\,0$.

\noindent The Hamiltonian in terms of the commutative variables $(x_i,p_i)$ then takes the form
\begin{equation}
H(t)=\dfrac{a(t)}{2}\left({p_1}^2+{p_2}^2\right)+\dfrac{b(t)}{2}\left({x_1}^2+{x_2}^2\right)+c(t)\left({p_1}{x_2}-{p_2}{x_1}\right)+d(t)\left(x_1p_1+p_1x_1+x_2p_2+p_2x_2\right)\,\,\,;\label{ham2}
\end{equation}
where $a(t), b(t), c(t), d(t)$ are the time dependent coefficients formed as
\begin{align}
&a(t)=\dfrac{1}{M}\left(1-\dfrac{\theta(t)\Omega(t)}{4} \right)+\dfrac{M\,\omega^2\theta^2(t)}{4}~,
b(t)=M\,\omega^2\left(1-\dfrac{\theta(t)\Omega(t)}{4} \right)+\dfrac{\Omega^2(t)}{4\,M}
~,\nonumber \\
&c(t)=\dfrac{\Omega(t)}{2M}+\dfrac{M\omega^2\theta(t)}{2}
~,~
d(t)= \dfrac{\sqrt{-\theta(t)\Omega(t)}}{4}\left[\dfrac{\Omega(t)}{2M}-\dfrac{M\omega^2\theta(t)}{2} \right]~.   \label{ham2co}              
\end{align}
Note that while the form of the Hamiltonian [eqn.(\ref{ham2})] and its time-dependent coefficients $a(t)$, $b(t)$, $c(t)$, and $d(t)$ [eqn.(\ref{ham2co})] are similar to those used in \cite{SG3} to investigate various explicit forms of geometric phases in NC space, our current study has different objectives from those in \cite{SG3}.

\section{Solution of the model Hamiltonian through Lewis-Riesenfeld approach}
We briefly review the Lewis-Riesenfeld approach \cite{Lewis}
to determine the eigenstate of the model Hamiltonian $H(t)$ [Eqn.(\ref{ham2})]. The approach involves constructing the time-dependent Hermitian invariant operator $I(t)$, also known as the Ermakov-Lewis invariant, that corresponds to the form of the Hamiltonian operator for our system.
According to this formalism, if we can determine the eigenstate of the operator $I(t)$,  
\begin{equation}
I(t)\ket{\phi}=\epsilon\,\ket{\phi}~;
\end{equation}
where $\epsilon$ denotes the time independent eigenvalue of $I(t)$ corresponding to its time dependent eigenstate $\ket{\phi}$, then one can also obtain the time dependent eigenstate of the operator $H(t)$, denoted by $\ket{\psi}$, using the following relation provided in \cite{Lewis},
\begin{equation}
\ket{\psi}=e^{i\Theta(t)}\ket{\phi}~.\label{L.eqn}
\end{equation}
The time dependent phase factor $\Theta(t)$,
commonly referred to as the Lewis phase, reads
\begin{equation}
\dot{\Theta}(t)=\bra{\phi}i\partial_t-H(t)\ket{\phi}~.\label{phase}
\end{equation}


\subsection{The Ermakov-Lewis invariant for our model system and its polar representation}
The initial step in solving the time-dependent model system using the method of invariants \cite{Lewis} is to obtain the Lewis invariant operator form of $I(t)$. The invariance of the operator $I(t)$ implies 
\begin{equation}
\dfrac{dI}{dt}=\partial_t{I}+\dfrac{1}{i}[I,H]=0~.
\end{equation}
As demonstrated in \cite{SG3}, the Lewis technique is utilized to solve the Hamiltonian described in eqn. (\ref{ham2}). Thus, for our present study, we adopt the final expression of $I(t)$ from \cite{SG3}. The invariant reads
\begin{equation}
I(t)=\rho^2({p_1}^2+{p_2}^2)+\left[\dfrac{\left(\dot{\rho}-2\rho\,d\right)^2}{a^2}+\dfrac{\xi^2}{\rho^2}\right]({x_1}^2+{x_2}^2)-\dfrac{\rho}{a}\left(\dot{\rho}-2\rho\,d\right)(x_1{p_1}+p_1{x_1}+x_2{p_2}+p_2{x_2})~ \label{inv}
\end{equation}
where $\rho=\rho(t)$, a time dependent parameter, obeys a non-linear differential equation given by
\begin{equation}
\ddot{\rho}-\dfrac{\dot{a}}{a}\dot{\rho}+\rho\,\left(ab-2\dot{d}-4d^2+2\dfrac{\dot{a}}{a}d\right)={\xi^2}\dfrac{a^2}{\rho^3}~.
\label{EP}
\end{equation}
In the subsequent discussion, the dimensionless constant of integration, denoted by $\xi$, will be set to $1$. This non-linear differential equation is commonly referred to as the Ermakov-Pinney (EP) equation in the literature. \\
\noindent 
For the sake of convenience in calculations, we present the Lewis invariant [Eqn.(\ref{inv})] in the form of polar coordinate variables. \footnote{Note that the polar coordinate variable $\theta$ should not be confused with the time-dependent NC parameter $\theta(t)$.} To achieve this, we adopt the same procedure as outlined in our previous communication~\cite{SG1}. The invariant in polar representation can be expressed as,
\begin{eqnarray}
I(t)=\dfrac{\xi^2}{\rho^2}r^2+\left(\rho{p_r}-\dfrac{\dot{\rho}-2\rho\,d}{a}r\right)^2+\left({\dfrac{\rho{p_\theta}}{r}}\right)^2-\left({\dfrac{\rho}{2r}}\right)^2~;
\label{invpol}
\end{eqnarray}
where the canonical coordinates in polar representation have the following 
form,
\begin{eqnarray}
p_r=-i\left({\partial}_r+\dfrac{1}{2r} \right)~~,~~
p_{\theta}=-i{\partial_{\theta}}~.
\label{polar}
\end{eqnarray}
It is relevant to mention that both the form of the hermitian invariant in Eqn.(\ref{invpol}) and EP equation in Eqn.(\ref{EP}), in the limit as $d(t)\rightarrow\,0$, revert to the same form as found in \cite{Dey}.
Here we also would like to point out that the time dependent Hermitian invariant operator $I(t)$, as obtained, is not unique. Instead, it is possible to get another alternative form of the invariant operator from the current expression of $I(t)$ [as given in Eqn.(\ref{inv})]. This alternative form will prove to be more advantageous in determining the eigenstate of the Hermitian invariant, and we will explore this further in the subsequent discussion.


\subsection{Eigenfunction associated with Ermakov-Lewis invariant}
To construct the eigenstate of a Hermitian invariant, we shall now obtain a more suitable form of the invariant operator using the expression in Eqn.(\ref{inv}). Once we have the new form of the invariant operator, we can proceed with the construction of the eigenstate. It is worth noting that this approach has also been utilized in \cite{Dey, SG3} to solve the invariant system.
\subsubsection{An alternative expression for the Lewis invariant}
Let us define a new form of invariant as,
\begin{equation}
I^{'}=\dfrac{I}{4}-\dfrac{p_\theta}{2}.\label{invn}
\end{equation}
It is very easy to verify that $I^{'}(t)$ also holds the property of invariance as $[p_\theta,H]=0$. Hence, in the discussion to follow, we will solve the newly defined form of the invariant introduced above to determine the eigenfunction of the Hamiltonian.
As analytically explained in \cite{Dey, SG3}, the choice of the alternative form as the final hermitian invariant to be solved is based on its ability to readily yield the following eigenvalue when solved with the appropriate form of the ladder operators,
\begin{eqnarray}
I^{'}\ket{n,l}=\left(n+\dfrac{1}{2}\right)\ket{n,l}.
\end{eqnarray}
\noindent It is obvious that $\ket{n, l}$ where $n$ and $l$ are the integers such that $n+l=m\,\geqslant0$, is an eigenstate of both $I^{'}$ and $p_{\theta}$ since $[I^{'}, p_{\theta}]=0$. Hence, for any given value of $n$ and $m$ which are non-negative integers, $l$ holds the 
condition $l=m-n$ with the range $l\in \lbrace -n,..,0,....\rbrace $.

\subsubsection{Eigenfunction of the alternative form of the invariant} 
As we have already mentioned, our Hamiltonian $H(t)$ has the same structure as the one presented in \cite{SG3}, and it was solved using the Lewis method of invariants. The detailed discussion of solving the invariant operator with the appropriate form of ladder operators is provided in \cite{SG3}. In our present work, we utilize the obtained expression of the eigenfunction of the invariant operator \cite{SG3}.


\noindent The eigenfunction of the invariant in the polar coordinate system reads \cite{SG3}.
\begin{eqnarray}
\phi_{n,m-n}(r,\theta, t)=\dfrac{i^m\,\rho^{m-n-1}}{\sqrt{m!n!\pi}}r^{n-m}e^{i(m-n)\theta-\dfrac{a-i\rho(\dot{\rho}-2\rho\,d)}{2a{\rho}^2}r^2}U\left(-m,1-m+n,\dfrac{r^2}{\rho^2} \right)~;\label{efU}
\end{eqnarray}
where $U\left(-m,1-m+n,\dfrac{r^2}{\rho^2} \right)$ is known as
Tricomi's confluent hypergeometric function \cite{Arfken, uva} which can also be expressed in terms of the 
associated Laguerre polynomial as
\begin{equation}
U\left(-m,1-m+n,\dfrac{r^2}{\rho^2} \right)=\dfrac{m!}{(-1)^m}\,L^{n-m}_m\left(\dfrac{r^2}{\rho^2} \right)~.
\end{equation}
Thus, the eigenfunction of the invariant operator can also be expressed as,
\begin{align}
\phi_{n,m-n}(r,\theta, t)&=\dfrac{i^{-m}\sqrt{m!}\,\rho^{m-n-1}}{\sqrt{n!\pi}}r^{n-m}e^{i(m-n)\theta-\dfrac{a-i\rho(\dot{\rho}-2\rho\,d)}{2a{\rho}^2}r^2}\,L^{n-m}_m\left(\dfrac{r^2}{\rho^2} \right)~\nonumber\\
&=Q_{n,m-n}(t)\,R_{n,m-n}(r,t)\,\Phi_{n,m-n}(\theta,t)\,;\label{efL}
\end{align}
where the functions $R_{n, m-n}(r,t), \Phi_{n, m-n}(\theta,t)$ and $Q_{n, m-n}(t)$ are given by 
\begin{align}
&R_{n,m-n}(r,t)=r^{n-m}e^{-\dfrac{a-i\rho(\dot{\rho}-2\rho\,d)}{2a{\rho}^2}r^2}\,L^{n-m}_m\left(\dfrac{r^2}{\rho^2} \right)
\,,\nonumber\\
&\Phi_{n,m-n}(\theta,t)=e^{i(m-n)\theta}~,~Q_{n,m-n}(t)=\dfrac{i^{-m}\sqrt{m!}\,\rho^{m-n-1}}{\sqrt{n!\pi}}~.\label{efLparts}
\end{align}
The eigenfunction $\phi_{n,m-n}(r,\theta,t)$ holds the following 
orthonormality condition,
\begin{equation}
\int_0^{2\pi}d\theta\int_0^{\infty}rdr\phi^{*}_{n,m-n}(r,\theta)\phi_{n^{'},m^{'}-n^{'}}(r,\theta)=\delta_{nn^{'}}\delta_{mm^{'}}.
\label{orth}
\end{equation}
\subsection{Lewis phase factor and the eigenfunction of the model Hamiltonian}
The Lewis phase factor's form, originally presented by Lewis {\it et al} in \cite{Lewis}, can be derived from the following condition [Eqn.(\ref{phase})], 
\begin{equation}
\dot{\Theta}(t)=\bra{\phi}i\partial_t-H(t)\ket{\phi}.
\end{equation}
The form was obtained in \cite{Dey, SG3} and reads,
\begin{equation}
\Theta_{n,m-n}(t)=\,m\,\int_0^{t}{\left[c(\tau)-\dfrac{a(\tau)}{\rho^2(\tau)}\right] d\tau}.\label{phase2}
\end{equation}
\noindent To construct the eigenfunction of the Hamiltonian Eqn.(\ref{ham2}), we now have all the ingredients, including Eqn(s).(\ref{efL}, \ref{phase2}). Using equation Eqn.(\ref{L.eqn}), the eigenfunction of H(t) can be written as follows

\begin{eqnarray}
\psi_{n,m-n}(r,\theta,t)&=&\dfrac{i^{-m}\sqrt{m!}\,\rho^{m-n-1}}{\sqrt{~n!\,\pi~}}\,exp\left[{i\,m\,\int_0^{t}{\left[c(\tau)-\dfrac{a(\tau)}{\rho^2(\tau)}\right] d\tau}}\right]\nonumber\\
&&\times\,r^{n-m}e^{i(m-n)\theta-\dfrac{a-i\rho(\dot{\rho}-2\rho\,d)}{2a{\rho}^2}r^2}L^{n-m}_m\left(\dfrac{r^2}{\rho^2} \right)~.\label{efLH}
\end{eqnarray}

\section{Exact analytical solutions of the Ermakov-Pinney equation}

In order to obtain the exact analytical solutions of the EP equation, we essentially follow~\cite{Dey}, where the exact analytical solutions of the EP equation [Eqn.(\ref{EP})] have been constructed for the case where $d(t)\,=\,0$ by following the Chiellini integrability condition \cite{chill}.

\noindent Here, we extend their solution set by obtaining the explicit value of the additional parameter $d(t)$ corresponding to the set of values of the parameters $a(t)$, $b(t)$, and $\rho(t)$ derived in \cite{Dey}.
it can then be verified that the parameters $a(t)$, $b(t)$, $\rho(t)$ and $d(t)$ also satisfy the Chiellini integrability condition.  

\subsection{Exponential Ermakov-Pinney solution} 
The exponential solution set of EP equation derived in \cite{Dey} is given by the following
relations, 
\begin{eqnarray} 
a(t)=\sigma e^{-\Gamma{t}}\,\,\,,\,\,\,b(t)=\Delta e^{\Gamma{t}}\,\,\,,\,\,\rho(t)={\mu}e^{-\Gamma/2}\,\,\,\,\,;\label{exp1}
\end{eqnarray}
where $\sigma,\Delta, \mu$ and $\Gamma$ (positive real number) are the constants constrained by the following relation

\begin{equation}
4\sigma\Delta\mu^4-\mu^4\Gamma^2-4\sigma^2=0~;\label{exp1c}
\end{equation}
which appears after the direct substitution of the above solution set [Eqn.(\ref{exp1})] into the form of the EP equation [Eqn.(\ref{EP})] at the limit $d(t)\,=\,0$.

\noindent  The expressions for $a(t), b(t)$ and $\rho(t)$ are now substituted into the EP equation (Eqn.[\ref{EP})]. This results in the emergence of the following differential equation that is obeyed by the additional parameter $d(t)$,
\begin{equation}
\dot{d}+2d^2+\Gamma\,d=\dfrac{4\mu^4\sigma\Delta-\mu^4\Gamma^2-4\sigma^2}{8\mu^4}\equiv\Bbbk\,\text{(constant)}~.\label{expk}
\end{equation}
The solution for $d(t)$ takes the form,
\begin{align}
d(t)\,=\,\dfrac{1}{4}\left[\sqrt{\Gamma^2+8\Bbbk}~~\dfrac{\complement\,e^{\sqrt{\Gamma^2+8\Bbbk}\,\,t}+1}{\complement\,e^{\sqrt{\Gamma^2+8\Bbbk}\,\,t}-1}-\Gamma \right]~~;\label{expd1}
\end{align}
where $\complement$, the integration constant, is set to be greater than unity to ensure that the obtained solution can no longer diverge at any time $t$. 
To elaborate the fact, we derive the critical time $t_0$ at which the above solution would diverge. The condition of divergence for the above solution is as follows,
\begin{equation}
\complement\,e^{\sqrt{\Gamma^2+8\Bbbk}\,\,t_0}-1=0
\end{equation}  
and the critical time $t_0$ found to be
\begin{equation}
t_0\,=\,-\dfrac{log\,\complement}{\sqrt{\Gamma^2+8\Bbbk}}~.
\end{equation}
Hence, by choosing the value of $\complement$ to be greater than $1$, we can make $t_0$ negative. This avoids the divergence, as physical time $t$ cannot approach $t_0$.

\subsubsection*{A special and convenient form of exponential EP solution\,:}
\noindent It is worth noting that Eqn.(\ref{expd1}) yields various simple forms of the solution when a suitable value of the constant $\Bbbk$ in Eqn.(\ref{expk}) is chosen. To understand this better,  we consider the simplest example by setting the value of $\Bbbk$ to zero. This results in the following relation as can be seen from Eqn.(\ref{expk}),
\begin{align}
\mu^4\sigma\Delta-\sigma^2=\dfrac{\mu^4\Gamma^2}{4}~;\label{expk0}
\end{align}  
which is basically identical to the relation in Eqn.(\ref{exp1c}).
\noindent Hence, the parameter $d(t)$ takes the form 
\begin{equation}
d(t)|_{\Bbbk=0}=\dfrac{\Gamma}{2\left(\complement\,e^{\Gamma\,t}-1\right)}~~;~~\complement\,>1~.\label{expd0}
\end{equation}

\noindent Note that $d(t)$ becomes zero in the limit $\complement\rightarrow\infty$. 

\subsubsection{Consistency with the Chilleni integrability condition}
In this subsection, we aim to demonstrate that these solution sets discussed earlier also satisfy the Chilleni integrability condition \cite{chill}, which was used in \cite{Dey} to derive the explicit expressions for the first three EP parameters, namely $a(t)$, $b(t)$, and $d(t)$. To accomplish this, we shall first write down the Chilleni condition below.

\noindent Defining certain components of the EP equation (Eqn. \ref{EP}) as follows
\begin{eqnarray}
\dot{\rho}=\eta~,~
g(\rho)=-\dfrac{\dot{a}}{a}~,~h(\rho)=\rho\left(ab-2\dot{d}-4\,d^2+2\dfrac{\dot{a}}{a}d \right)-\dfrac{a^2}{\rho^3}~;\label{chil1}
\end{eqnarray}
enables to transform Eqn.(\ref{EP}) into the following first order differential equation 

\begin{equation}
\eta\dfrac{d\eta}{d\rho}+\eta\,g(\rho)+h(\rho)=0~.\label{chil2}
\end{equation}
Now the Chilleni integrability condition states that if
\begin{align}
\dfrac{d}{d\rho}\left(\dfrac{h(\rho)}{g(\rho)}\right)=q\,g(\rho)~;~(q=\text{constant})\label{chil3}
\end{align}
then the unknown parameter $\eta$ has the following solution as \cite{chill}
\begin{align}
\eta=\lambda_q\dfrac{h(\rho)}{g(\rho)}~~~\text{with}~~~
\lambda_q=\dfrac{-1\pm\sqrt{1-4\,q}}{2\,q}~.\label{chil4}
\end{align}
Substituting the expressions for exponential EP solution set $a(t), b(t), \rho(t)$ and $d(t)$ together with the constraint relation [Eqn.(\ref{expk})] in Eqn.(\ref{chil1}), the explicit form of $\eta(\rho), g(\rho)$ and $h(\rho)$ are calculated to be
\begin{align}
\eta(\rho)=-\dfrac{\Gamma}{2}\rho~~,~~g(\rho)=\Gamma~~,~~
h(\rho)=\dfrac{\Gamma^2}{4}\rho~.\label{chil5}
\end{align}
With the above set of values in our hand, the Chilleni integrability condition stated in Eqn(s).(\ref{chil3}, \ref{chil4}) can be easily shown to be satisfied. The corresponding values of the constants $q$ and $\lambda_q$, for this solution set, are computed to be 
\begin{eqnarray}
q=\dfrac{1}{4}~~,~~\lambda_q=-2~;
\end{eqnarray}
which also obey the relation among $q$ and $\lambda_q$.

\subsection{Rational Ermakov-Pinney solution}

The rational solution set of EP equation derived in \cite{Dey} is given by the following
relations, 
\begin{eqnarray}
a(t)=\dfrac{\sigma\,\left(1+\dfrac{2}{k}\right)^{\,(k+2)/k}}{(\Gamma{t}+\chi)^{\,(k+2)/k}}~,~
b(t)=\dfrac{\Delta\,\left(\dfrac{k}{k+2} \right)^{(2-k)/k} }{(\Gamma{t}+\chi)^{\,(k-2)/k}}~,~\rho(t)=\dfrac{\mu\left(1+\dfrac{2}{k}\right)^{1/k} }{(\Gamma{t}+\chi)^{1/k}}~;\label{rat1}
\end{eqnarray}
where $k\,\,\in\,\,N$ and $\sigma, \Delta, \mu, \chi$ and $\Gamma$ (positive real number) are the constants constrained by the following relation
\begin{equation}
\left(k+2\right)^2\left(\sigma\Delta\mu^4-\sigma^2 \right)-\mu^4\Gamma^2=0~;\label{ratc1}
\end{equation}
which appears after the direct substitution of the above solution set [Eqn.(\ref{rat1})] in the reduced form of the EP equation [Eqn.(\ref{EP})] with $d=\,0$.  \\
\noindent 
Now we substitute the expression of $a(t),b(t) \,$and$\, \rho(t)$ in the EP equation [Eqn.(\ref{EP})]. This leads to the following differential equation for the additional parameter $d(t)$,
\begin{align}
\dfrac{1}{k^2\left(\Gamma{t}+\chi\right)^{2}}\left[\left(k+2\right)^2\left(\sigma\Delta\mu-\dfrac{\sigma^2}{\mu^3}\right)-\mu\Gamma^2\right]=2\mu\left[\dot{d}(t)+2d^2(t)+d(t)\dfrac{\Gamma(k+2)}{k(\Gamma\,t+\chi)} \right]~.\label{ratdiff}
\end{align}
It can be seen that the time dependent factors in the above equation would cancel if the unknown parameter $d(t)$ has the following form   
\begin{equation}
d(t)=\dfrac{\delta}{\left(\Gamma\,t+\chi\right)}~;\label{ratd}
\end{equation}  
where $\delta$ is a positive, real constant. With the above form of $d(t)$, Eqn.(\ref{ratdiff}) now reduces to a 
constraint relation of the form, 
\begin{equation}
4k^2\mu^4\delta\left(\delta+\dfrac{\Gamma}{k} \right)=\left(k+2\right)^2\left(\sigma\Delta\mu^4-\sigma^2 \right)-\mu^4\Gamma^2~.\label{ratc2}
\end{equation}
Hence, the EP solution in terms of rational functions are given by Eqn(s).(\ref{rat1}, \ref{ratd}) with the constraint relation Eqn.(\ref{ratc2}).
It should be noted that while the parameters $a(t)$, $b(t)$, and $\rho(t)$ take on various special forms due to different values of $k$, the parameter $d(t)$ remains entirely independent of $k$.

\subsubsection{Consistency with the Chilleni integrability condition}
In this subsection, we show that the rational solution sets satisfy the Chilleni integrability condition discussed in subsection $4.1.1$. To do so, we substitute the expressions for the rational EP solution set, namely, $a(t), b(t), \rho(t),$ and $d(t),$ along with the constraint relation [Eqn.(\ref{ratc2})], into Eqn.(\ref{chil1}). By doing this, we obtain explicit expressions for $\eta(\rho),$ $g(\rho),$ and $h(\rho)$,
\begin{align}
\eta(\rho)=-\dfrac{\rho^{k+1}\Gamma}{\mu^k(k+2)}~~;~~g(\rho)=\Gamma\dfrac{\rho^k}{\mu^k}~~;~~
h(\rho)=\dfrac{\rho^{2k+1}\Gamma^2}{\mu^{2k}\left(k+2\right)^2}~.\label{chil5rat}
\end{align}
With the above set of values in our hand, the Chilleni integrability condition stated in Eqn(s).(\ref{chil3}, \ref{chil4}) can easily be verified and the corresponding values of the constants $q$ and $\lambda_q$ are given by 
\begin{eqnarray}
q=\dfrac{k+1}{\left(k+2\right)^2}~~,~~\lambda_q=-\left(k+2\right)~;
\end{eqnarray}
which also obey the relation $\lambda_q=(-1-\sqrt{1-4\,q})\,/2\,q$ among $q$ and $\lambda$.

\section{Expectation values and uncertainty relations}
We now aim to obtain the expression for energy expectation value, and the uncertainty relations among the coordinate and momentum operators. 


\noindent The expression for $\langle H \rangle$, arising from Eqn[\ref{ham2}], is given by
\begin{equation}
\langle H\rangle = \dfrac{a(t)}{2}\left(\langle{p_1}^2\rangle+\langle{p_2}^2\rangle\right)+\dfrac{b(t)}{2}\left(\langle{x_1}^2\rangle+\langle{x_2}^2\rangle\right)+c(t)\left(\langle{p_1}{x_2}-{p_2}{x_1}\rangle\right)+d(t)\left(\langle{x_1}{p_1}+{p_1}{x_1}\rangle+\langle{x_2}{p_2}+{p_2}{x_2}\rangle\right)\,\,\,.\label{enrgexp}
\end{equation}
Note that the eigenstates of the Hamiltonian $H(t)$ are denoted by $|n,l\rangle_H$. 

\subsection{Expectation values of coordinate operators, bilinear products, and the operators raised to the second power} 

\noindent To begin with, we present the expectation values of the configuration space operators $x_1$ and $x_2$, as well as their respective squared forms, $x_1^2$ and $x_2^2$. It is noteworthy that the computation of these quantities, when raised to any arbitrary finite power, has been previously elaborated in \cite{SG1} where we investigated a damped harmonic oscillator in a time-dependent background of noncommutativity. 
The expectation values can be written as follows \cite{Dey, SG3},
\begin{align}
_{H}\langle n,m-n|\,x_i\,|n,m-n\rangle_{H}\,&=\,0~,\nonumber\\
_{H}\langle n,m-n|\,x_i^2\,|n,m-n\rangle_{H}\,&=\,\dfrac{(m+n+1)}{2}\,\rho^2~;\label{xsq}
\end{align}
where $i=1, 2$. 
Our results, obtained using a generalized version of the Bopp-shift relations, are consistent with those from \cite{Dey, SG1}, where a time-dependent NC framework was developed using standard Bopp-Shift relations. It happens due to the fact that, in our generalized eigenfunction (Eqn.[\ref{efLH}]) , the extra parameter $d(t)$ generated due to the modification included in the coordinate mapping relations, exists as a phase. 
As a result, the factor $d(t)$ does not contribute since it is eliminated during the calculation of expectation value.

\noindent 
We will now calculate the expectation values of the momentum operators, $p_1$ and $p_2$, as well as their respective squared forms, $p_1^2$ and $p_2^2$. These operators can be expressed in polar form as follows,
\begin{eqnarray}
p_1\,=\,\left[-i\,cos\,\theta\,\partial_r+\dfrac{i}{r}sin\theta\,\partial_{\theta}\right]~,~p_2\,=\,\left[-i\,sin\,\theta\,\partial_r-\dfrac{i}{r}cos\theta\,\partial_{\theta}\right]~.
\end{eqnarray}
Since the detailed calculation reveals that both $\braket{p_1}$ and $\braket{p_2}$ yield trivial zero results, we just present those in the following manner
\begin{equation}
_{H}\langle n,m-n|\,p_i\,|n,m-n\rangle_{H}\,=\,0\,;~~i\,=\,[1,2]\,,\label{pi}
\end{equation}
and then proceed to calculate $\braket{\,p_1^2\,}$ and $\braket{\,p_2^2\,}$. 
\noindent We mention some of the intermediate calculational steps in getting the final result. The expectation value of $p_1^2$ reads

\begin{eqnarray} 
_{H}\langle n,m-n|\,p_1^2\,|n,m-n\rangle_{H}&=&\int r dr d\theta
~_{H}\langle n,m-n |r,\theta\rangle\langle r,\theta|\left(-i\,cos\,\theta\,\partial_r+\dfrac{i}{r}sin\theta\,\partial_{\theta}\right)^2|n,m-n\rangle_{H}\nonumber\\
&=&\int r dr d\theta
~\langle n,m-n |r,\theta\rangle\langle r,\theta|\left(-i\,cos\,\theta\,\partial_r+\dfrac{i}{r}sin\theta\,\partial_{\theta}\right)^2|n,m-n\rangle\nonumber\\
&=&\int
~\phi_{n,m-n}^{*}(r,\theta,t)\left(-i\,cos\,\theta\,\partial_r+\dfrac{i}{r}sin\theta\,\partial_{\theta}\right)^2\phi_{n,m-n}(r,\theta,t)\, r dr d\theta~;
\nonumber\\
&=&\,I_1+ I_2+ I_3+ I_4+ I_5 ;\label{psq1}
\end{eqnarray} 
where we have used the relations 
$|n,m-n\rangle_{H}~=~e^{i\Theta_{n,m-n}}|n,m-n\rangle$ where $|n,m-n\rangle_H$ and 
$|n,m-n\rangle$ denote the eigenstates of the Hamiltonian $H(t)$ and Lewis invariant $I(t)$ 
respectively. We have also employed the relation $\langle r,\theta|n,m-n\rangle$~=~$\phi_{n,m-n}(r,\theta)$, with $\phi$ presenting the eigenfunction of $I(t)$. 
We now calculate the terms $I_1, I_2, I_3, I_4$ and $I_5 $. 
\begin{eqnarray} 
_{H}\langle n,m-n|\,p_1^2\,|n,m-n\rangle_{H}=\,I_1+ I_2+ I_3+ I_4+ I_5 ;\label{psq2}
\end{eqnarray} 
\begin{align}
I_1=&-\int\,\phi_{n,m-n}^{*}(r,\theta,t)\,cos^2\theta\,\partial_r^2\,\left[\phi_{n,m-n}(r,\theta, t)\right]\,r\,dr\,d\theta \nonumber
\\
&=-Q^{*}_{n,m-n}(t)\,Q_{n,m-n}(t)\,\int_0^\infty\,R_{n,m-n}^{*}(r,t)\,\left[\partial_r^2\,R_{n,m-n}(r,t)\right]\,r\,dr\,\int_0^{2\Pi}\,\Phi_{n,m-n}^{*}(\theta,t)\,cos^2\theta\,\Phi_{n,m-n}(\theta,t)\,d\theta~,\label{I1}
\end{align}
\begin{align}
I_2&=-2\int\,\phi_{n,m-n}^{*}(r,\theta, t)\,\left(\dfrac{sin\,\theta\,cos\theta}{r^2}\right)\,\partial_\theta\,\left[\phi_{n,m-n}(r,\theta, t)\right]\,r\,dr\,d\theta 
\nonumber
\\
&=-2\,Q^{*}_{n,m-n}(t)\,Q_{n,m-n}(t)\,\int_0^\infty\,\dfrac{R_{n,m-n}^{*}(r, t)\,\,R_{n,m-n}(r, t)}{r}\,dr\,\int_0^{2\Pi}\,\Phi_{n,m-n}^{*}(\theta, t)\,cos\theta\,sin\theta\,\partial_{\theta}\left[\Phi_{n,m-n}(\theta, t)\right]\,d\theta~,\label{I2}
\end{align}
\begin{align}
I_3&=2\int\,\phi_{n,m-n}^{*}(r,\theta, t)\,\left(\dfrac{sin\,\theta\,cos\theta}{r}\right)\,\partial_r\partial_\theta\,\left[\phi_{n,m-n}(r,\theta, t)\right]\,r\,dr\,d\theta 
\nonumber
\\
&=2\,Q^{*}_{n,m-n}(t)\,Q_{n,m-n}(t)\,\int_0^\infty\,R_{n,m-n}^{*}(r, t)\,\partial_r\,\left[R_{n,m-n}(r, t)\,\right]dr\,\int_0^{2\Pi}\,\Phi_{n,m-n}^{*}(\theta, t)\,cos\theta\,sin\theta\,\partial_{\theta}\left[\Phi_{n,m-n}(\theta, t)\right]\,d\theta~,\label{I3}
\end{align}
\noindent
\begin{align}
I_4&=-\int\,\phi_{n,m-n}^{*}(r,\theta, t)\,\left(\dfrac{sin^2\theta}{r}\right)\,\partial_r\,\left[\phi_{n,m-n}(r,\theta, t)\right]\,r\,dr\,d\theta\nonumber\\
&=-Q^{*}_{n,m-n}(t)\,Q_{n,m-n}(t)\,\int_0^\infty\,R_{n,m-n}^{*}(r, t)\,\left[\partial_r\,R_{n,m-n}(r, t)\right]\,dr\,\int_0^{2\Pi}\,\Phi_{n,m-n}^{*}(\theta, t)\,sin^2\theta\,\Phi_{n,m-n}(\theta, t)\,d\theta~, \label{I4}
\end{align}
\begin{align}
I_5&=-\int\,\phi_{n,m-n}^{*}(r,\theta, t)\,\left(\dfrac{sin^2\theta}{r^2}\right)\,\partial_\theta^2\,\left[\phi_{n,m-n}(r,\theta, t)\right]\,r\,dr\,d\theta\nonumber\\
&=-Q^{*}_{n,m-n}(t)\,Q_{n,m-n}(t)\,\int_0^\infty\,\dfrac{R_{n,m-n}^{*}(r, t)\,R_{n,m-n}(r, t)}{r}\,dr\,\int_0^{2\Pi}\,\Phi_{n,m-n}^{*}(\theta, t)\,sin^2\theta\,\partial_\theta^2\,\left[\Phi_{n,m-n}(\theta, t)\right]\,d\theta~.\label{I5}
\end{align}
\noindent Upon observing the azimuthal portions of the above integrals in Eqn(s).(\ref{I1}, \ref{I2}, \ref{I3}, \ref{I4}, \ref{I5}) , it becomes clear that the terms $I_2$ and $I_3$ vanish. This can be seen as
\begin{align}
&\int_0^{2\Pi}\,\Phi_{n,m-n}^{*}(\theta, t)\,cos\theta\,sin\theta\,\partial_{\theta}\left[\Phi_{n,m-n}(\theta, t)\right]\,d\theta\nonumber\\
&=\,i\,(m-n)\,\int_0^{2\Pi}\,\Phi_{n,m-n}^{*}(\theta, t)\,cos\theta\,sin\theta\,\Phi_{n,m-n}(\theta, t)\,d\theta\nonumber\\
&=0~.\label{psq00}
\end{align}
As a result, we have $I_2\,=\,0\,=I_3$.
To proceed further, we insert the expressions for $R$, $\Phi$, and $Q$ from Eqn. [\ref{efLparts}] into the remaining variables $I_1$, $I_4$, and $I_5$, and subsequently introduce the variable $z$ to parametrize $\dfrac{r^2}{\rho^2}$. In order to simplify the integrations, we utilize numerous valuable and specific relationships such as the derivative form, an identity relation, and the orthonormality condition of the associated Laguerre polynomial. The derivative of the associated Laguerre polynomial reads \cite{Arfken},
\begin{align}
\partial_z\left[L^{n-m}_m\,\left(z\right)\right]&=
-\,L^{n-m+1}_{m-1}\,\left(z\right)~;~(m\,\geq\,1)\nonumber\\
&=\,0~~~~~~~~~~~~~~~~~~~;(m=0)~.\nonumber\\
\end{align}
The orthonormality condition and the identity relation for the associated Laguerre polynomial reads \cite{Arfken}
\begin{align}
\int_0^\infty\,z^{n-m}\,e^{-z}\,L^{n-m}_{n}(z)\,L^{n-m}_{m}(z)
\,=\,\dfrac{\Gamma\,(2n-m+1)}{n!}\,\delta_{mn}~,\nonumber\\
L^{n-m}_{m}(z)\,=\,L^{n-m+1}_{m}(z)-L^{n-m+1}_{m-1}(z)~.\label{psqoi}
\end{align}
Using the above relations (Eqn.[\ref{psqoi}]) we can calculate the expressions $I_1, I_4$, and $I_5$ which read 
\begin{align}
I_1&=-\dfrac{m!\,(n-m)(n-m-1)}{2\,n!\,\rho^2}\,\int^{\infty}_0\,z^{n-m-1}\,e^{-z}\,L^{n-m}_{m}(z)L^{n-m}_{m}(z)\,dz\,+\,(2n+2m+1)\,\dfrac{f}{2\rho}-\dfrac{f^2}{2}(m+n+1)\nonumber\\
&+(2n-2m+1)\dfrac{m!}{\rho^2\,n!}\,\int_0^{\infty}z^{n-m}\,e^{-z}\,L^{n-m}_{m}(z)L^{n-m+1}_{m-1}(z)\,dz-\dfrac{2\,m!}{n!\rho^2}\,\int_0^{\infty}\,z^{n-m+1}\,e^{-z}\,L^{n-m}_{m}(z)L^{n-m+2}_{m-2}(z)\,dz, 
\end{align}

\begin{align}
I_4=-\dfrac{m!\,(n-m)}{2\,n!\,\rho^2}\,\int^{\infty}_0\,z^{n-m-1}\,e^{-z}\,L^{n-m}_{m}(z)L^{n-m}_{m}(z)\,dz\,+\,\dfrac{f}{2\rho}
+\dfrac{m!}{\rho^2\,n!}\,\int_0^{\infty}z^{n-m}\,e^{-z}\,L^{n-m}_{m}(z)L^{n-m+1}_{m-1}(z)\,dz~,
\end{align}
\begin{align}
I_5=\dfrac{m!\,(m-n)^2}{2\,n!\,\rho^2}\,\int^{\infty}_0\,z^{n-m-1}\,e^{-z}\,L^{n-m}_{m}(z)L^{n-m}_{m}(z)\,dz~
\end{align}
where $f=\dfrac{a-i\rho(\dot{\rho}-2\rho\,d)}{a\rho}$. Summing them up,

\begin{align}
_{H}\langle n,m-n|\,p_1^2\,|n,m-n\rangle_{H}&=\,\dfrac{2(n-m+1)\,m!}{n!\,\rho^2}\,\int_0^\infty\,z^{n-m}\,e^{-z}\,L^{n-m}_m(z)\,L^{n-m+1}_{m-1}(z)\,dz\nonumber\\
&-\,\dfrac{2\,m!}{n!\,\rho^2}\,\int_0^\infty\,z^{n-m+1}\,e^{-z}\,L^{n-m}_m(z)\,L^{n-m+2}_{m-2}(z)\,dz\nonumber\\
&+\dfrac{(n+m+1)}{2}\,\left[\dfrac{1}{\rho^2}+\dfrac{(\dot{\rho}-2\rho\,d)^2}{a^2} \right]~.\label{psqincom}
\end{align}
Now we use the following identity (derived in the Appendix A)
\begin{equation}
(n-m+1)\,\,\int_0^\infty\,z^{n-m}\,e^{-z}\,L^{n-m}_m(z)\,L^{n-m+1}_{m-1}(z)\,dz=\,\int_0^\infty\,z^{n-m+1}\,e^{-z}\,L^{n-m}_m(z)\,L^{n-m+2}_{m-2}(z)\,dz~.
\end{equation}
Using the above identity, Eqn.(\ref{psqincom}) reduces to the form
\begin{align}
_{H}\langle n,m-n|\,p_1^2\,|n,m-n\rangle_{H}\,=\,\dfrac{(n+m+1)}{2}\,\left[\dfrac{1}{\rho^2}+\dfrac{(\dot{\rho}-2\rho\,d)^2}{a^2} \right]~;\label{psq3}
\end{align}
which is a real and positive quantity. The presence of the parameter $d(t)$ in the above expression enables us to visualize the effect of employing the generalized coordinate mapping relation between NC space and commutative space. When the parameter $d(t)$ vanishes, our result reduces to that previously obtained in \cite{Dey}. Furthermore, in our system, the appearance of $d(t)$ is solely due to presence of NC parameters, allowing us to identify its presence as a pure NC effect in the above expression.

\noindent 
We can follow a similar procedure to obtain the expectation value of $p_2^2$., which yields the same result as $\braket{p_1^2}$. For this we have,
\begin{align}
_{H}\langle n,m-n|\,p_2^2\,|n,m-n\rangle_{H}\,=\,\dfrac{(n+m+1)}{2}\,\left[\dfrac{1}{\rho^2}+\dfrac{(\dot{\rho}-2\rho\,d)^2}{a^2} \right]~.\label{psq4}
\end{align}
Eqn.(s)(\ref{psq3}, \ref{psq3}) can be expressed in the following manner,
\begin{equation}
_{H}\langle n,m-n|\,p_i^2\,|n,m-n\rangle_{H}\,=\,\dfrac{(n+m+1)}{2}\,\left[\dfrac{1}{\rho^2}+\dfrac{(\dot{\rho}-2\rho\,d)^2}{a^2} \right]\,;~~i\,=\,[1, 2]\,.\label{psqi}
\end{equation}
Our next goal is to determine the expectation value of $\left(x_1\,p_1+p_1\,x_1\right)$. To accomplish this, we must first obtain the polar form of the said quantity. This reads
\begin{align}
\left(x_1\,p_1+p_1\,x_1\right)=2\,x_1\,p_1\,-i&\nonumber\\
=\,-2i\,\left(\,r\,cos^2\theta\,\partial_r-cos\theta\,sin\theta\,\partial_\theta\,\right)-i&~.
\end{align}
We shall now employ the same methodology utilized in computing $\braket{p_i^2}$. This now gives,
\begin{align} 
&_{H}\langle n,m-n|\left(x_1\,p_1+p_1\,x_1\right)|n,m-n\rangle_{H}\nonumber\\
&=\int r dr d\theta
~_{H}\langle n,m-n |r,\theta\rangle\langle r,\theta|\left[\,-2i\,\left(\,r\,cos^2\theta\,\partial_r-cos\theta\,sin\theta\,\partial_\theta\,\right)-i\right]|n,m-n\rangle_{H}\nonumber\\
&=\int r dr d\theta
~\langle n,m-n |r,\theta\rangle\langle r,\theta|\,\left[\,-2i\,\left(\,r\,cos^2\theta\,\partial_r-cos\theta\,sin\theta\,\partial_\theta\,\right)-i\right]|n,m-n\rangle\nonumber\\
&=-i-2i\int
~\phi_{n,m-n}^{*}(r,\theta, t)\,\left(\,r\,cos^2\theta\,\partial_r-cos\theta\,sin\theta\,\partial_\theta\,\right)\phi_{n,m-n}(r,\theta, t)\, r dr d\theta\nonumber\\
&=\,-i+I_1^{'}+I_2^{'}~;
\end{align} 
where 
\begin{align}
I_1^{'}=-2i\,Q^{*}_{n,m-n}\,Q_{n,m-n}\,\int_0^\infty\,R_{n,m-n}^{*}(r, t)\,r^2\,\left[\partial_r\,R_{n,m-n}(r, t)\right]\,dr\,\int_0^{2\Pi}\,\Phi_{n,m-n}^{*}(\theta, t)\,cos^2\theta\,\Phi_{n,m-n}(\theta, t)\,d\theta
\end{align}
\begin{align}
I_2^{'}=2i\,Q^{*}_{n,m-n}(t)\,Q_{n,m-n}(t)\,\int_0^\infty\,R_{n,m-n}^{*}(r, t)\,R_{n,m-n}(r, t)\,r\,dr\,\int_0^{2\Pi}\,\Phi_{n,m-n}^{*}(\theta, t)\,cos\theta\,sin\theta\,\partial_\theta\left[\Phi_{n,m-n}(\theta, t)\right]\,d\theta~.
\end{align}
Referring back to Eqn.(\ref{psq00}), we can conclude that 
\begin{equation}
I_2^{'}=0~.
\end{equation}
We proceed to evaluate the integral $I_1'$ using Eqn. (\ref{psqoi}), which leads us to the final result of $\langle\,x_1p_1+p_1x_1\rangle$ as,

\begin{align}
_{H}\langle n,m-n|\left(x_1\,p_1+p_1\,x_1\right)|n,m-n\rangle_{H}=(m+n+1)\,\dfrac{\rho\,\left(\dot{\rho}-2\rho\,d\right)}{a}~.\label{si1}
\end{align}

\noindent A similar calculation reveals that the quantity $\langle x_2p_2+p_2x_2\rangle$ also yields the result that we obtained in Eqn.(\ref{si1}). Therefore, we can represent it separately below, 
\begin{align}
_{H}\langle n,m-n|\left(x_2\,p_2+p_2\,x_2\right)|n,m-n\rangle_{H}=(m+n+1)\,\dfrac{\rho\,\left(\dot{\rho}-2\rho\,d\right)}{a}~.\label{si2}
\end{align}
or alternatively, present both of them in the following manner
\begin{align}
_{H}\langle n,m-n|\left(x_i\,p_i+p_i\,x_i\right)|n,m-n\rangle_{H}=(m+n+1)\,\dfrac{\rho\,\left(\dot{\rho}-2\rho\,d\right)}{a}~;\,i=1, 2\,.\label{si}
\end{align}

\noindent The last quantity that remains to be calculated is $\braket{H(t)}$ is $\braket{x_2\,p_1-p_2\,x_1}$, which in terms of polar coordinates reads
\begin{equation}
\left(x_2\,p_1-p_2\,x_1\right)\,=\,i\partial_{\theta}~.
\end{equation} 
 
The corresponding expectation value which can be derived much more simply than those we elaborated for the previous quantities, is mentioned below as follows,  
\begin{align}
_{H}\langle n,m-n|\left(x_2\,p_1-p_2\,x_1\right)|n,m-n\rangle_{H}=(n-m)~;\label{zee}
\end{align}
while each part of the term $\braket{x_2\,p_1-p_2\,x_1}$ individually reads
\begin{align}
_{H}\langle n,m-n|\left(x_1\,p_2\right)|n,m-n\rangle_{H}=\dfrac{(m-n)}{2}~;~_{H}\langle n,m-n|\left(x_2\,p_1\right)|n,m-n\rangle_{H}=\dfrac{(n-m)}{2}~.\label{zee1}
\end{align} 

\noindent Now that we have all the necessary results to compute $\braket{H(t)}$, we proceed to calculate the expectation values of two more bilinear operators, namely, $x_1x_2$ and $p_1p_2$. The expectation values of these operators are required to compute the uncertainty relations for the NC variables which we shall discuss later.
\begin{align}
x_1x_2\,=&\,r^2\,sin\theta\,cos\theta~,\nonumber\\
p_1\,p_2\,=\,-sin\theta\,cos\theta\,\partial_r^2+\dfrac{cos\,2\theta}{r^2}\,\partial_{\theta}-&\dfrac{cos\,2\theta}{r}\partial_r\partial_{\theta}+\dfrac{sin\theta\,\cos\theta}{r}\partial_r+\dfrac{sin\theta\,cos\theta}{r^2}\partial_{\theta}^2~.
\end{align}
Carefully observing the azimuthal parts of these quantities, with the experiences from the previous calculations in mind, we can easily recognize that neither of these quantities make any contribution to their expectation value in the eigenstate of $H(t)$. 
Hence, we mention the following outcomes,
\begin{align}
\braket{x_1\,x_2}=0~,~\braket{p_1\,p_2}=0 ~.\label{bi}
\end{align}

\subsection{Energy expectation values: Explicit form and their graphical investigation}
This section is dedicated to explore the time evolution of the energy expectation value of the system, utilizing both analytical and graphical representations. In the previous section, we have already performed the necessary calculations to compute the expectation value of the Hamiltonian $H(t)$ in its own eigenstate $\psi_{n,m-n}(r, \theta)$. Hence, the expectation value of energy $\braket{E_{n,m-n}(t)}$ with 
respect to energy eigenstate $\psi_{n,m-n}(r,\theta,t)$ can be derived from Eqn.(\ref{enrgexp}) as
\begin{align}
\braket{E_{n,m-n}(t)}&=\dfrac{1}{2}\,(n+m+1)\left[b(t)\rho^2(t)+\dfrac{a(t)}{\rho^2(t)}+\dfrac{\dot{\rho}^2(t)-4\rho^2(t)\,d^2(t)}{a(t)} \right]+c(t)\,(n-m)~;\label{enr}
\end{align}
which in the limit $d(t)\rightarrow\,0,$ gives the expectation value obtained in \cite{SG1}. 

\noindent  In a previous section discussing the solution of the EP equation, we derived the explicit functional form of the Hamiltonian parameter as well as the EP parameter $d(t)$ based on the previously calculated EP parameters $a(t), b(t)$, and $\rho(t)$, which were presented in \cite{Dey}. Now we aim to compute the explicit form of the energy expression in Eqn.(\ref{enr}) by substituting these various solution sets of the EP equation. Although the Hamiltonian parameter $c(t)$ in the energy expression above is not considered an EP parameter, it can still be explicitly calculated, as demonstrated in \cite{SG1}. By substituting the explicit form of the parameters, appearing in the Hamiltonian, as well as the EP parameters $a(t)$, $b(t)$, and $d(t)$, into their respective relations shown in Eqn.(\ref{ham2co}), one can obtain the explicit expressions of the NC parameters $\theta(t)\Omega(t)$, which in turn reveal the explicit form of the parameter $c(t)$ for constant value of the mass $M$ and the angular frequency $\omega_0$. To simplify things, we proceed with Eqn.(\ref{enr}) $n=m$. This gives 
\begin{align}
\braket{E_{m,0}(t)}&=\dfrac{1}{2}\,(2\,m+1)\left[b(t)\rho^2(t)+\dfrac{a(t)}{\rho^2(t)}+\dfrac{\dot{\rho}^2(t)-4\rho^2(t)\,d^2(t)}{a(t)} \right]~.\label{enr1}
\end{align}

\subsubsection{Energy expectation values : exponential EP solutions}
\begin{figure}[H]
\centering
\includegraphics[scale=1]{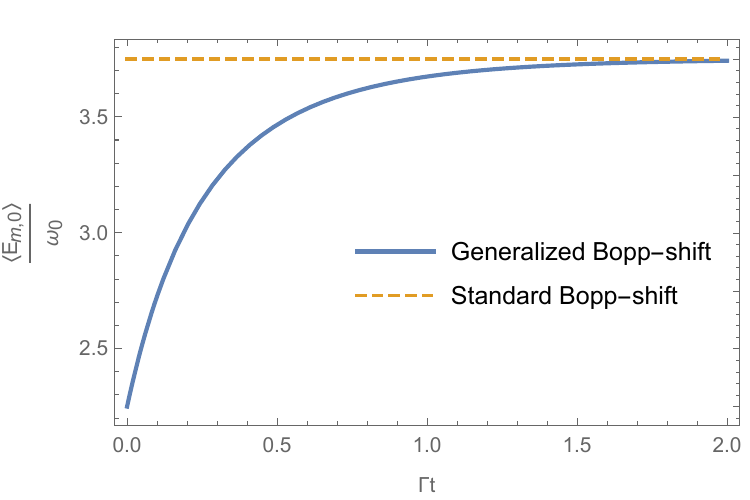}
\caption{\textit{Plot of the variation of expectation value of energy, (scaled by 
$\frac{1}{\omega_0}$ ($\frac{\langle E_{m,0} \rangle}{\omega_0}$) in order to make 
it dimensionless), as we vary $\Gamma$t. The expectation value of 
energy $\langle E \rangle$ is calculated for exponentially varying 
Hamiltonian parameters as well as EP parameters when $\langle A\rangle$ the system is mapped by 
the generalized version of the Bopp-shift relations and the parameters are considered (as per Eqn.(\ref{exp1c})) as $m, n, \mu, \Gamma=1$ and $\Delta=\dfrac{5}{4},\,\complement=2$; $\langle B\rangle$ the system is mapped by 
the standard Bopp-shift relations and the parameters are considered (as per Eqn.(\ref{exp1c})) as $m, \mu, \omega_0, \sigma, \Gamma=1$, $\Delta=\dfrac{5}{4}$ and $\complement\,\rightarrow\,\infty$.  While 
for $\langle A\rangle$ the energy exhibits an initial increase, followed by a saturation, 
for $\langle B\rangle$ the energy value always remains constant. }}  
\label{fig1}
\end{figure}

Considering the exponentially varying solution expressed in Eqn.(\ref{exp1}) and Eqn.(\ref{expd0}) subject to the constraint relation in Eqn.(\ref{expk0}), we can derive the expression for the energy expectation value, with $n=m$, as follows
\begin{align}
\braket{\,H(t)\,}|_{m=n}^{exp}\,=\,(2m+1)\left[\Delta\,\mu^2-\dfrac{\Gamma^2\,\mu^2}{2\,\sigma\,\left(\complement\,e^{\Gamma\,t}-1\right)^2}   \right]~;(\,\complement\,>\,1)~.\label{Eexp}
\end{align}
By employing the limit $\complement\rightarrow\infty$, the above expression reduces to a constant value, which represents the energy associated with the exponential EP solution in \cite{SG1}, while considering $n=m$. Both the analytic expression (Eqn. [\ref{Eexp}]) and its graphical representation (see Fig. [\ref{fig1}]) reveal that for real, positive values of the constants which are chosen as per Eqn.(\ref{exp1c}) for both the cases, the energy increases with time and eventually reaches a finite value at large time. Here, the graphical representation also nicely reflects the fact that, at any instant, due to the deduction of energy in amount $4\rho^2d^2$ (as outlined in Eqn.(\ref{enr})), the energy associated with the generalized Bopp-shift transformation will not exceed the same associated with the standard Bopp-shift transformation.

\subsubsection{Energy expectation values : rational EP solutions}
Considering the rationally varying solution expressed in Eqn.(\ref{rat1}) and Eqn.(\ref{ratd}) subject to the constraint relation in Eqn.(\ref{ratc2}), we can derive the expression for the energy expectation value, with $n=m$, as follows

\begin{align}
\braket{\,H(t)\,}|_{m=n}^{rat}\,=\,\dfrac{(2m+1)}{\left(\Gamma\,t+\chi \right)}\left[\dfrac{(k+2)^2\,\left(\sigma^2+\Delta\mu^4\sigma\right)+\Gamma^2\mu^4-4\mu^4\delta^2k^2}{2\,(k+2)\,\mu^2\,k\,\sigma\,}\right]~.\label{Erat}
\end{align}
In the limit $\delta\rightarrow\,0$ the above expression represents the energy that corresponds to the rational EP solution in \cite{SG1}, when $n=m$ is taken into consideration. Although the energy decays rationally over time in both cases, the coefficients of the decay are altered due to the removal of the modified term from the coordinate mapping relation (Eqn.[\ref{Pbop}]). In contrast to the observations in the previous example (Fig. 1), the graphical representation indicates that, at any given point, the energy associated with the generalized Bopp shift transformation surpasses that of the standard Bopp shift transformation. This discrepancy arises from the fact that the constants associated with the rational solution set adhere to the constraint relation in Eqn. (\ref{ratc2}) for the generalized system, whereas they conform to the constraint relation in Eqn. (\ref{ratc1}) for the standard system. 
\begin{figure}[h!]
\centering
\includegraphics[scale=1]{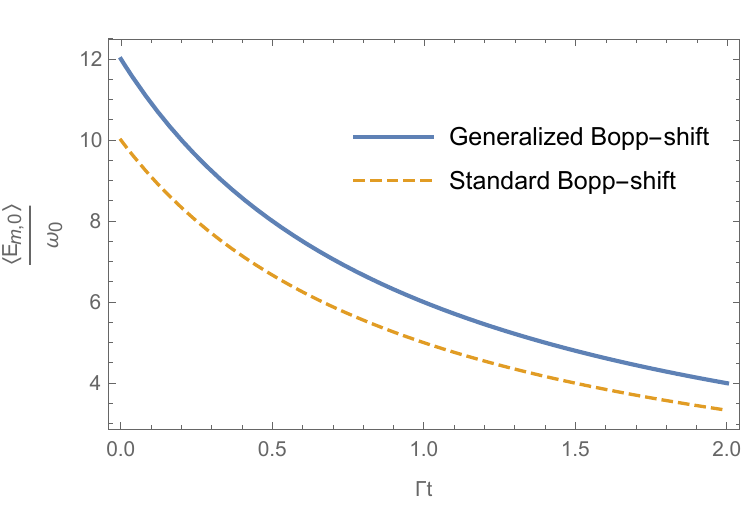}
\caption{\textit{Plot of the variation of expectation value of energy, (scaled by 
$\frac{1}{\omega_0}$ ($\frac{\langle E_{m,0} \rangle}{\omega_0}$) in order to make 
it dimensionless), as we vary $\Gamma$t. The expectation value of 
energy $\langle E \rangle$ is  calculated for rationally varying 
Hamiltonian parameters as well as EP parameters when $\langle A\rangle$ the system is mapped by 
the generalized version of the Bopp-shift relations and the parameters are considered (as per Eqn.(\ref{ratc2})) as $m, n, k, \mu, \Gamma, \omega_0,  \delta, \sigma, \chi=1$ and $\Delta=2$; $\langle B\rangle$ the system is mapped by 
the standard Bopp-shift relations and the parameters are considered (as per Eqn.(\ref{ratc1})) as $m, k, \mu, \Gamma, \sigma, \chi=1$ and $\Delta=\dfrac{10}{9}$.  In both cases, the energy, containing different values of constant coefficients, decays rationally with respect to time and approaches zero for very large values of time.}}  
\label{fig2}
\end{figure}

\subsection{Modified uncertainty equality relations for commutative and noncommutative variables}
In this section, our objective is to find uncertainty relations between the operators. 

\subsubsection*{Equal uncertainty products for commutative operators :}
The variances for the variables $x_1$ and $x_2$ are obtained from Eqn.(\ref{xsq}) as 
\begin{align}
\Delta\,x_1=\sqrt{\braket{x_1^2}\,-\,\braket{x_1}^2}\,=\,\dfrac{\sqrt{m+n+1}}{\sqrt{2}}\,\rho~,~
~\Delta\,x_2=\sqrt{\braket{x_2^2}\,-\,\braket{x_2}^2}\,=\,\dfrac{\sqrt{m+n+1}}{\sqrt{2}}\,\rho ~;
\end{align}
and the same for the variables $p_1$ and $p_2$ are deduced from Eqn(s).(\ref{pi}, \ref{psqi}) as 
\begin{align}
\Delta\,p_1=\sqrt{\braket{p_1^2}\,-\,\braket{p_1}^2}\,=\dfrac{\sqrt{m+n+1}}{a\,\rho}\sqrt{\dfrac{a^2+\rho^2\left(\dot{\rho}-2\rho\,d\right)^2}{2}}~;\nonumber\\
\Delta\,p_2=\sqrt{\braket{p_2^2}\,-\,\braket{p_2}^2}\,=\,\dfrac{\sqrt{m+n+1}}{a\,\rho}\sqrt{\dfrac{a^2+\rho^2\left(\dot{\rho}-2\rho\,d\right)^2}{2}}~. 
\end{align}
The equal uncertainty products among the parameters $x_i$ and $p_i$, where $i\,=\,[1,2]$, are therefore given by,
\begin{align}
\Delta\,x_1\,\Delta\,y_1=\,\left(m+n+1\right)\dfrac{\rho^2}{2}~&,~
\Delta\,p_1\,\Delta\,p_2=\dfrac{(n+m+1)}{2}\left[\dfrac{1}{\rho^2}+\dfrac{(\dot{\rho}-2\rho\,d)^2}{a^2} \right]\,~,\nonumber\\
\Delta\,x_1\,\Delta\,p_1=\Delta\,x_2\,\Delta\,p_2&=\dfrac{m+n+1}{2\,a}\sqrt{a^2+\rho^2\left(\dot{\rho}-2\rho\,d\right)^2}~.
\end{align}
The uncertainties in the product mentioned above yield the values reported in \cite{Dey} as the parameter $d(t)$ tends to zero. 

\subsubsection*{Equal uncertainty products for noncommutative operators :}

\noindent
Organizing the quantities derived in Eqn(s).(\ref{xsq}, \ref{pi}) based on the coordinate mapping relation presented in Eqn.(\ref{Pbop}), we can determine that 
\begin{align}
\braket{X_i}=0~;~\braket{P_i}=0~~~;~~~i\in [1,2]~.
\end{align}   
Now doing the same thing for the quantities derived in Eqn(s).(\ref{xsq}, \ref{psqi}, \ref{si}, \ref{zee1}, \ref{bi}) we establish that
\begin{align}
\braket{X_i^2}=\dfrac{(m+n+1)}{2}\,\left[\rho^2\,\left(1-\dfrac{\theta\,\Omega}{4}\right)+\dfrac{\theta^2}{4}\,\left(\dfrac{1}{\rho^2}+\dfrac{\left(\dot{\rho}-2\rho\,d\right)^2}{a^2}\right) 
-\dfrac{\theta\sqrt{-\theta\,\Omega\,}\rho}{2\,a}\left(\dot{\rho}-2\rho\,d \right)  \right]-(m-n)\,\dfrac{\theta}{2}~,
\end{align}
\begin{align}
\braket{P_i^2}=\dfrac{(m+n+1)}{2}\,\left[\,\left(1-\dfrac{\theta\,\Omega}{4}\right)\left(\dfrac{1}{\rho^2}+\dfrac{(\dot{\rho}-2\rho\,d)^2}{a^2}\right)+\dfrac{\Omega^2\rho^2}{4}\,
+\dfrac{\Omega\sqrt{-\theta\,\Omega\,}\rho}{2\,a}
\left(\dot{\rho}-2\rho\,d \right)  \right]-(m-n)\,\dfrac{\Omega}{2}~.
\end{align}
With these values in hand, we now proceed to determine the product of variances for the NC operators. This leads to  
\begin{align}
&\Delta\,X\,\Delta\,Y=\sqrt{\braket{X^2}\braket{Y}^2}=\braket{X^2}=\braket{Y^2}\nonumber\\
&=\dfrac{(m+n+1)}{2}\,\left[\rho^2\,\left(1-\dfrac{\theta\,\Omega}{4}\right)+\dfrac{\theta^2}{4}\,\left(\dfrac{1}{\rho^2}+\dfrac{(\dot{\rho}-2\rho\,d)^2}{a^2}\right) 
-\dfrac{\theta\sqrt{-\theta\,\Omega\,}\rho}{2\,a}\left(\dot{\rho}-2\rho\,d \right)  \right]-(m-n)\,\dfrac{\theta}{2}~,
\end{align}
\begin{align}
&\Delta\,P_X\,\Delta\,P_Y=\sqrt{\braket{P_X^2}\braket{P_Y}^2}=\braket{P_X^2}=\braket{P_Y^2}\nonumber\\
&=\dfrac{(m+n+1)}{2}\left[\,\left(1-\dfrac{\theta\,\Omega}{4}\right)\left(\dfrac{1}{\rho^2}+\dfrac{(\dot{\rho}-2\rho\,d)^2}{a^2}\right)+\dfrac{\Omega^2\rho^2}{4}
+\dfrac{\Omega\sqrt{-\theta\Omega\,}\rho}{2\,a}
\left(\dot{\rho}-2\rho\,d \right)  \right]-(m-n)\dfrac{\Omega}{2}~,
\end{align}
\begin{align}
&\Delta\,X\,\Delta\,P_X=\Delta\,Y\,\Delta\,P_Y=\sqrt{\braket{X^2}\braket{P_X^2}}=\sqrt{\braket{Y^2}\braket{P_Y^2}}\nonumber\\
&=\left\lbrace\dfrac{(m+n+1)}{2}\,\left[\rho^2\,\left(1-\dfrac{\theta\,\Omega}{4}\right)+\dfrac{\theta^2}{4}\,\left(\dfrac{1}{\rho^2}+\dfrac{\left(\dot{\rho}-2\rho\,d\right)^2}{a^2}\right) 
-\dfrac{\theta\sqrt{-\theta\,\Omega\,}\rho}{2\,a}\left(\dot{\rho}-2\rho\,d \right)  \right]-(m-n)\,\dfrac{\theta}{2} \right\rbrace^{\dfrac{1}{2}}\nonumber\\
&\times\left\lbrace\dfrac{(m+n+1)}{2}\,\left[\,\left(1-\dfrac{\theta\,\Omega}{4}\right)\left(\dfrac{1}{\rho^2}+\dfrac{(\dot{\rho}-2\rho\,d)^2}{a^2}\right)+\dfrac{\Omega^2\rho^2}{4}\,
+\dfrac{\Omega\sqrt{-\theta\,\Omega\,}\rho}{2\,a}
\left(\dot{\rho}-2\rho\,d \right)  \right]-(m-n)\,\dfrac{\Omega}{2}\right\rbrace^{\dfrac{1}{2}}~.
\end{align}
It is nice to note that our results are consistent with those found in \cite{Dey} as the parameter $d$ and the multiplication of the NC parameters $\theta\Omega$ tend to zero.

\section{Conclusions}
In this work, our main focus is to investigate the behaviour of a NC harmonic oscillator in time dependent background, as studied in \cite{Dey}, when it is mapped in terms of commutative variables using a generalized version of the standard Bopp-shift relations, recently introduced by \cite{spb}. The primary motivation for this study is to look for exact solutions of some prototype quantum mechanical systems in time dependent NC backgrounds. Such a study would enrich the understanding of such systems. Specifically, we aim to investigate how this approach modifies the behaviours of the oscillator undergoing mapping via the standard Bopp-shift relations in \cite{Dey}. For doing so, a two-dimensional harmonic oscillator is considered in NC framework with time-dependent NC parameters. To facilitate our analysis, we employ the generalized version of the standard Bopp-shift relation \cite{spb} to transform the system into commutative variables. The generalized version of the Bopp shift relating the NC variables to the commutative variables have resulted in getting a geometric phase shift in planar NC quantum mechanics. Hence, an interesting question that arises is that whether such kind of transformations allow exact analytical solutions of quantum systems. This is another motivation for the present work. We then present the exact eigenfunction expression of the Hamiltonian, previously derived in \cite{SG3}, which addresses a similar time-dependent system using the Lewis invariant associated with the non-linear Ermakov-Pinney (EP) equation. Considering the potential benefits of solving the EP equation, which acts as a constraint relation among the Hamiltonian parameters, we proceed to explicitly solve this equation. As our work is an extended version of the system described in \cite{Dey}, where the exact analytical solutions of the EP equation were found by following the Chiellini integrability condition \cite{chill}, we expand upon their solution set. We obtained the explicit value of the additional EP parameter that corresponds to the set of values of the EP parameters derived in \cite{Dey}. Here we observed some interesting features about the newly derived EP parameter both in exponential and rational form. In the exponential solution set, while the first three EP parameters, found in \cite{Dey}, have distinct individual form, the fourth one appears with a generic structure. This structure allows for a special form of the solution to be derived by setting a suitable value for a certain constant present in the general form of the parameter. We also provided an example illustrating the emergence of this special functional form for the fourth EP parameter. In the case of rationally varying EP solution set, the power of variation for the first three EP parameters, as provided in \cite{Dey}, is not fixed, as it depends on a positive valued, real constant. However, the fourth EP parameter just varies inversely with respect to time. We then verify that both the EP solution set in expanded form, which varies exponentially and rationally with respect to time, is consistent with the Chiellini integrability condition. Next, we compute the expectation value of the Hamiltonian in a generic form and show that, in the absence of the modified term from the coordinate mapping relations, it reproduces the result obtained in our previous communication \cite{SG1}, where we used the standard Bopp shift relations to map a system of damped harmonic oscillator from NC space to commutative space. The generic form of the energy expectation value in the eigenstates of the Hamiltonian, for a particular choice of quantum number, is explicitly explored through the explicit exponential and rational form of the EP solution set. In addition to creating a graphical representation of the explicit energy expression for both the rationally varying and exponentially varying cases, we also compared each case graphically with its behaviour associated with the standard Bopp-shift relations. Initially, the dynamics of the exponential energy expression shows an increase with increasing time and then saturates. Throughout this process, the associated behaviour governed by the standard Bopp shift relations remains constant over time. In contrast, the dynamics of the rational energy expression, containing different coefficients of variations, gradually decrease to zero over time. It can be concluded regarding the energy dynamics of the system that, if it is possible to uniformly select the explicit form of the time dependent parameters and the numerical value of the constants for both cases, the modified version of the Bopp-shift relation, at any given instant, would not permit the energy of the system to exceed that associated with the standard Bopp-shift relation. In the final part of our work, we computed variances for both commutative and NC operators, and used these values to determine the generalized form of uncertainty equality relations between them. Such a study is important in its own sight as it would lead to a deeper understanding of the quantum theory in such time dependent NC backgrounds. Expectedly, the equality relations reproduce the results obtained in \cite{Dey} in appropriate limits. Although the inequality relations would complete the analysis, we would like to report them in future work. Before we end, we would like to mention that at present the experimental status of noncommutativity is far fetched as observing Planck scale signatures is beyond the scope of present technology. 

\section*{Acknowledgement}
MD would like to thank Mr. Soham Sen for having useful discussion with him and for his helpful assistance in operating the software Mathematica.

\section*{Appendix A: Associated Laguerre polynomial identity}
\noindent In this Appendix, we shall prove the following identity involving the associated Laguerre polynomial. 
\begin{align}
(n-m+1)\,\int_0^\infty\,z^{n-m}\,e^{-z}\,L^{n-m}_m(z)\,L^{n-m+1}_{m-1}(z)\,dz\,-\,\int_0^\infty\,z^{n-m+1}\,e^{-z}\,L^{n-m}_m(z)\,L^{n-m+2}_{m-2}(z)\,dz=0~.\label{ap1}
\end{align}
We begin by describing another identity related to the associated Laguerre polynomial. The identity reads
\begin{align}
\,z\,L^{n-m+1}_{m-1}(z)\,=\,n\,L^{n-m}_{m-1}(z)-\,m\,L^{n-m}_{m}(z)~.\label{ap2}
\end{align}
We can express the left-hand side of Eqn.(\ref{ap1}) in the following manner,
\begin{align}
(n-m+1)\,\int_0^\infty\,z^{n-m}\,e^{-z}\,L^{n-m}_m(z)\,L^{n-m+1}_{m-1}(z)\,dz\,-\,\int_0^\infty\,z^{n-m}\,e^{-z}\,L^{n-m}_m(z)\,\left[z\,L^{n-m+2}_{m-2}(z)\right]\,dz=0~.\label{ap3}
\end{align}
As per the identity in Eqn.(\ref{ap2}), 
\begin{align}
z\,L^{n-m+2}_{m-2}(z)\,=\,n\,L^{n-m+1}_{m-2}(z)-\,(m-1)\,L^{n-m+1}_{m-1}(z)~.\label{ap4}
\end{align}
By substituting the above identity into Eqn.(\ref{ap3}), and performing some algebraic calculations, we obtain
\begin{align}
(n-m+1)\,\int_0^\infty\,z^{n-m}\,e^{-z}\,L^{n-m}_m(z)\,L^{n-m+1}_{m-1}(z)\,dz-\,\int_0^\infty\,z^{n-m}\,e^{-z}\,L^{n-m}_m(z)\,\left[z\,L^{n-m+2}_{m-2}(z)\right]\,dz\nonumber\\
=n\,\int_0^\infty\,z^{n-m}\,e^{-z}\,L^{n-m}_{m}(z)\,\left[\,L^{n-m+1}_{m-1}(z)-L^{n-m+1}_{m-2}(z)\,\right]\,dz~.
\end{align}

\noindent Applying the second relation in Eqn.(\ref{psqoi}), the expression above can be simplified as,
\begin{align}
n\,\int_0^\infty\,z^{n-m}\,e^{-z}\,L^{n-m}_{m}(z)\,\left[\,L^{n-m+1}_{m-1}(z)-L^{n-m+1}_{m-2}(z)\,\right]\,dz
=n\,\int_0^\infty\,z^{n-m}\,e^{-z}\,L^{n-m}_m(z)\,L^{n-m}_{m-1}(z)\,dz~.
\end{align}
Now the right hand side of the above relation turns out to be zero as per the orthonormality condition mentioned in Eqn.(\ref{psqoi}). Therefore, we have
\begin{align}
(n-m+1)\,\int_0^\infty\,z^{n-m}\,e^{-z}\,L^{n-m}_m(z)\,L^{n-m+1}_{m-1}(z)\,dz\,&-\,\int_0^\infty\,z^{n-m+1}\,e^{-z}\,L^{n-m}_m(z)\,L^{n-m+2}_{m-2}(z)\,dz\nonumber\\
&=n\,\int_0^\infty\,z^{n-m}\,e^{-z}\,L^{n-m}_m(z)\,L^{n-m}_{m-1}(z)\,dz\nonumber\\
&=0~.
\end{align}



\end{document}